\documentclass[12pt,fleqn]{article}
\usepackage[DIV12]{typearea}
\usepackage{amsmath,amsfonts,amssymb}
\usepackage{graphicx}
\usepackage{cite}
\usepackage{mathrsfs}
\usepackage{mathtools}
\usepackage{bbm}
\usepackage{braket}
\usepackage{pifont}
\usepackage{units}
\usepackage[small,loose]{subfigure}  

\newcommand{\Eqref}[1]{equation~\eqref{#1}}
\newcommand{\Figref}[1]{figure~\ref{#1}}

\newcommand{\Secref}[1]{section~\ref{#1}}

\newcommand{\eVdist}{\kern-0.06em}

\newcommand{\CenterObject}[1]{\ensuremath{\vcenter{\hbox{#1}}}}

\newcommand{\I}{\mathrm{i}}

\newcommand{\E}[1]{\ensuremath{\mathrm{E}_{#1}}} 
\newcommand{\G}[1]{\ensuremath{\mathrm{G}_{#1}}}
\newcommand{\SO}[1]{\ensuremath{\mathrm{SO}(#1)}}
\newcommand{\SU}[1]{\ensuremath{\mathrm{SU}(#1)}}

\newcommand{\Z}[1]{\ensuremath{\mathbbm{Z}_{#1}}} 
\newcommand{\V}[0]{\ensuremath{\boldsymbol{V}}}

\newcommand{\Id}{\ensuremath{\mathbbm{1}}}

\allowdisplaybreaks[1]

\def\MyTitle{A note on discrete $\boldsymbol{R}$ symmetries\\[0.2cm] in $\boldsymbol{\Z{6}}$--II orbifolds with Wilson lines}
\title{\MyTitle}

\begin{document}

\begin{titlepage}

\begin{flushright}
\normalsize{DESY-13-143}\\
\normalsize{TUM-HEP 901/13}\\
\normalsize{FLAVOUR-EU 52/13}\\
\end{flushright}

\vspace*{1.0cm}

\begin{center}
{\Large\bf\MyTitle
}

\vspace{1cm}

\textbf{
Hans~Peter Nilles\footnote[1]{Email: \texttt{nilles@th.physik.uni-bonn.de}}{}$^a$,
Sa\'ul~Ramos--S\'anchez\footnote[2]{Email: \texttt{ramos@fisica.unam.mx}}{}$^b$,
Michael~Ratz\footnote[3]{Email: \texttt{michael.ratz@tum.de}}{}$^c$,
Patrick~K.S.~Vaudrevange\footnote[4]{Email: \texttt{patrick.vaudrevange@desy.de}}{}$^d$
}
\\[5mm]
\textit{\small
{}$^a$ Bethe Center for Theoretical Physics\\
{\footnotesize and}\\
Physikalisches Institut der Universit\"at Bonn,\\
Nussallee~12, 53115~Bonn, Germany
}
\\[3mm]
\textit{\small
{}$^b$ Department of Theoretical Physics, Physics Institute, UNAM\\ 
~~Mexico D.F.\ 04510, Mexico
}
\\[3mm]
\textit{\small
{}$^c$ Physik-Department T30, Technische Universit\"at M\"unchen, \\
~~James-Franck-Stra\ss e, 85748~Garching, Germany
}
\\[3mm]
\textit{\small
{}$^d$ Deutsches Elektronen-Synchrotron DESY, \\
~~Notkestra\ss e  85, 22607~Hamburg, Germany
}
\end{center}

\vspace{1cm}

\begin{abstract}
We re--derive the
$R$ symmetries for the \Z6--II orbifold with non--trivial Wilson
lines and find expressions for the $R$ charges which differ from those in the
literature.
\end{abstract}

\end{titlepage}

\section{Introduction}

$R$ symmetries play a key role in understanding supersymmetric field theories
and in model building. It is well known that $R$ symmetries do arise from the
Lorentz symmetry of compact dimensions. In many cases the compact dimensions
only have discrete isometries, leading to discrete $R$ symmetries in the
effective four--dimensional (4D) theory. This is, in particular, true for
orbifold compactifications \cite{Dixon:1985jw,Dixon:1986jc}.

In the past, $R$ symmetries have been derived for the case of \Z6--II orbifold
compactifications of the heterotic string \cite{Kobayashi:2004ya}. Later it was
observed in \cite{Araki:2008ek} that, unlike all other continuous and discrete
symmetries of the effective 4D description of these settings, the
$\mathbbm{Z}_M^R$ symmetries have non--universal anomalies. This already
suggested that there might be something wrong with the $R$ charges. And, 
indeed, more recently it was pointed out in \cite{Bizet:2013gf}
that the $R$ charges have to be amended by contributions from so--called
$\gamma$ phases. The purpose of this note is to re--derive the $R$ symmetries
and charges for the \Z6--II orbifold, and to clarify the situation. Moreover,
our re--derivation allows us to determine the $R$ charges also in settings with
non--trivial Wilson lines.

This note is organized as follows. Section~\ref{sec:symmetries} contains our
re--derivation of $R$ symmetries and charges. Finally, 
section~\ref{sec:summary} contains our conclusions, including a brief 
discussion of the implications of the correct charges for model
building.

\section{Discrete $\boldsymbol{R}$ symmetries in $\boldsymbol{\Z6}$--II orbifolds}
\label{sec:symmetries}

After a brief introduction to the \Z6--II orbifold in \Secref{sec:Z6IIOrbifold} 
we discuss the origin of discrete $R$ symmetries in \Secref{sec:Sublattice}. In
\Secref{sec:GeometricalEigenstate} we derive previously unknown contributions
to the $R$ charges, which turn out to be essential in order to make the
corresponding discrete anomalies universal, such that they can be cancelled
by the dilaton via the Green--Schwarz mechanism.

\subsection{The $\boldsymbol{\Z{6}}$--II orbifold}
\label{sec:Z6IIOrbifold}

\begin{figure}[!h!]
\begin{center}
\hspace*{-0.5cm}
\subfigure[$\mathbbm{T}_{\G2}^2$.]{\CenterObject{\includegraphics{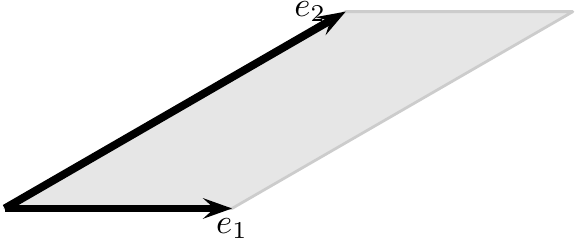}}}
~
\subfigure[$\mathbbm{T}_{\SU3}^2$.\label{fig:SU3Torus}]{\CenterObject{\includegraphics{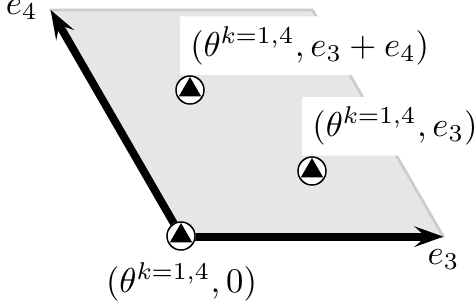}}}
~
\subfigure[$\mathbbm{T}_{\SU2\times\SU2}^2$.\label{fig:SO4Torus}]{\CenterObject{\includegraphics{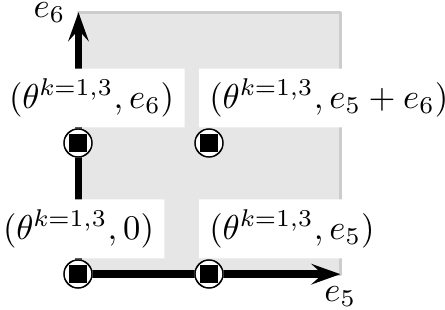}}}
\hspace*{-0.5cm}
\end{center}
\caption{The factorized lattice $\Lambda$ of the six--torus is chosen to be 
spanned by the roots of $\text{G}_2\times\SU{3}\times\SU{2}^2$. The six vectors 
$e_\alpha$ denote a basis. For later convenience, we show also the fixed points
in the \SU3 and $\SU2\times\SU2$ planes.}
\label{fig:Z6lattice}
\end{figure}

The \Z6--II orbifold is defined as the quotient space of the six--dimensional
torus  $\mathbbm{T}^6$ by the point group $P = \Z{6}$,
\begin{equation}
\label{eqn:Z6IIOrbifold}
\mathbbm{O} ~=~ \mathbbm{T}^6/P ~=~ \mathbbm{C}^3/\mathbbm{S}\;.
\end{equation}
The generator of $\Z{6}$ is denoted as $\theta$ with $\theta^6=\Id$. For 
\Z6--II it is represented by the so--called twist vector
\begin{equation}
\label{eqn:twist}
v~=~\left(0,\frac{1}{6},\frac{1}{3},-\frac{1}{2}\right)\;,
\end{equation}
which specifies the rotational angles as fractions of $2\pi$ in the three 
complex planes, i.e.\ the three complex torus--coordinates $z^i$ get mapped to
$\mathrm{e}^{2\pi\,\I\,v^i}\,z^i$ for $i=1,2,3$ and $v^0 = 0$ for later convenience. 
The twist acts on the factorized six--torus
$\mathbbm{T}^6=\mathbbm{T}_{\G2}^2\times\mathbbm{T}_{\SU3}^2\times\mathbbm{T}^2_{\SU2\times\SU2}$
(see \Figref{fig:Z6lattice}), whose defining  six--dimensional lattice $\Lambda$
is given by the root lattice of  $\text{G}_2\times\SU{3}\times\SU{2}^2$. 

Equivalently, one can define the orbifold $\mathbbm{O}$ as the quotient space of
$\mathbbm{C}^3$ by the so--called space group $\mathbbm{S}$, see 
\Eqref{eqn:Z6IIOrbifold}. Elements of $\mathbbm{S}$ are of the form  $g =
\left(\theta^k, n_\alpha\, e_\alpha\right)$ with summation over
$\alpha=1,\ldots,6$,  $k = 0,\ldots,5$, $n_\alpha \in \Z{}$ and $e_\alpha$
denote six basis vectors of the torus--lattice $\Lambda$.  $g$ acts on $z
\in \mathbbm{C}^3$ as  $z \mapsto g\,z ~=~ \theta^k\, z +
n_\alpha\, e_\alpha$ and the equivalence relation 
\begin{equation}
 z ~\sim~ g\,z \quad\text{for}\quad g \in \mathbbm{S} \quad\text{and}\quad 
 z \in \mathbbm{C}^3
\end{equation}
defines the orbifold. For a consistent compactification of the heterotic string 
on $\mathbbm{O}$ one has to embed the action of $g \in \mathbbm{S}$ into the 16 
gauge degrees of freedom of $\E{8}\times\E{8}$ or $\SO{32}$, which we denote  by
$X^I$ with $I=1,\ldots,16$: the twist $\theta$ acts as a shift $V^I$ and 
lattice translations by $e_\alpha$ are accompanied by Wilson lines
$W_\alpha^I$,  both restricted by modular invariance.  $g$ acts
simultaneously on $z$ and $X$ as
\begin{equation}
z ~\xmapsto{g}~ \theta^k\, z + n_\alpha\, e_\alpha 
\quad\text{and}\quad X ~\xmapsto{g}~ X 
+\pi\,\left(k\, V + n_\alpha\, W_\alpha\right)\;.
\end{equation}
As usual, one associates to $g = \left(\theta^k,
n_\alpha\, e_\alpha\right)$ the local twist $v_g = k\, v$ and the local shift 
$V_g = k\, V + n_\alpha\, W_\alpha$.

Consider a massless, closed (twisted) string with boundary condition given by 
$g = \left(\theta^k, n_\alpha\, e_\alpha\right) \in \mathbbm{S}$, i.e.\ 
$\boldsymbol{Z}(\tau,\sigma + \pi) = g\, \boldsymbol{Z}(\tau, \sigma)$ for the 
three complex world--sheet bosons $\boldsymbol{Z}$ on $\mathbbm{O}$. After 
canonical quantization this string can be described schematically by a state of 
the form 
\begin{equation}
\label{eq:twistedstate}
\Ket{p_\mathrm{sh},q_\mathrm{sh},\widetilde{N},\widetilde{N}^*,g}
~=~
\Ket{q_\text{sh}}_\text{R} \otimes 
\left(\widetilde{\alpha}^{i}_{-\omega_i}\right)^{\widetilde{N}^i}\, 
\left(\widetilde{\alpha}^{\bar{\imath}}_{-1+\omega_i}\right)^{\widetilde{N}^{*i}} \Ket{p_\text{sh}}_\text{L} 
\otimes \Ket{g}\;,
\end{equation}
where R and L denote the right-- and left--movers with shifted momenta 
$q_\text{sh}$ and $p_\text{sh}$, respectively. Here $q_\text{sh}=q+v_g$ 
with $q$ from either the vectorial or spinorial weight lattice of $\SO{8}$, and 
$p_\text{sh}=p+V_g$ with $p$ from the $\E{8}\times\E{8}$ weight lattice. We use 
the convention that the number of $-\nicefrac{1}{2}$ in the spinorial weight 
lattice is even. Then, $q_\text{sh}(\text{boson}) = q_\text{sh}(\text{fermion}) + 
(\nicefrac{1}{2},-\nicefrac{1}{2},-\nicefrac{1}{2},-\nicefrac{1}{2})$. As usual, 
fermions with $q_\text{sh}^0 = -\nicefrac{1}{2}$ are left--chiral. Further, 
$\Ket{g}$ specifies the localization of the string as follows. If $k \neq 0$, a 
string twisted by $g = \left(\theta^k, n_\alpha\, e_\alpha\right)$ is  localized 
at some fixed point or fixed torus $f_g \in \mathbbm{C}^3$, i.e.\   $g\, f_g =
f_g$ with $f_g$ being the coordinates of the fixed point or fixed  torus. We 
will refer to $g$ as ``constructing element'' for the corresponding  massless 
mode. Furthermore, the left--moving ground state  $\Ket{p_\text{sh}}_\text{L}$ 
can be excited by oscillators: in each (complex) direction $i=1,2,3$ and 
$\bar{\imath}=\bar{1},\bar{2},\bar{3}$ there are $\widetilde{N}^i$ excitations 
with $\widetilde{\alpha}^{i}_{-\omega_i}$ and  $\widetilde{N}^{*i}$ 
excitations with $\widetilde{\alpha}^{\bar{\imath}}_{-1+\omega_i}$. In the $-1$ 
ghost picture, this state is created by the vertex operator
\begin{equation}\label{eq:V-1}
 \V^{(g)}_{\!\!-1}~=~\mathrm{e}^{-\boldsymbol{\phi}}\, 
 \mathrm{e}^{2\I\,q_\mathrm{sh}\cdot \boldsymbol{H}}\,
 \mathrm{e}^{2\I\,p_\mathrm{sh}\cdot \boldsymbol{X}}\,
 \prod_{i=1}^3 \left(\partial \boldsymbol{Z}^i\right)^{\widetilde{N}^{i}}\,
 \left(\partial \boldsymbol{Z}^{*\,i}\right)^{\widetilde{N}^{*i}}\,
 \boldsymbol{\sigma}_{g}\;.
\end{equation}
In particular, the state $\Ket{g}$ is created by the twist field
$\boldsymbol{\sigma}_{g}$.

Selection rules are derived from correlators of vertex
operators~\cite{Hamidi:1986vh,Dixon:1986qv},
\begin{equation}\label{eq:cor}
 \mathcal{A}~=~\left\langle \V^{(g_1)}_{\!\!-1/2}\,\V^{(g_2)}_{\!\!-1/2}\,
 \V^{(g_3)}_{\!\!-1}\,\V^{(g_4)}_{\!\!0}
 \ldots \V^{(g_{L})}_{\!\!0}\right\rangle \;.
\end{equation}
The correlation function \eqref{eq:cor} factorizes into correlators involving
separately the fields $\boldsymbol{\phi}$, $\boldsymbol{X}^I$, 
$\boldsymbol{\sigma_{g}}$, $\boldsymbol{H}$ and $\boldsymbol{Z}^i$ 
\cite{Hamidi:1986vh,Dixon:1986qv,Font:1988tp,Font:1988mm,Font:1989aj}. This
leads to the condition of gauge invariance, the so--called space group selection
rules and to discrete $R$ symmetries as we explain in what follows.

\subsection{Discrete $\boldsymbol{R}$ symmetries and sublattice rotations}
\label{sec:Sublattice}

Discrete $R$ symmetries are intimately connected with so--called sublattice 
rotations. Since $\mathbbm{O}$ is factorized, $\mathbbm{O}$ respects 
symmetries beyond the elements of $\mathbbm{S}$, given by the sublattice 
rotations $\theta^{(j)}$ for $j=1,2,3$, i.e.\ separate rotations in each 
two--torus, corresponding to the three twist vectors
\begin{equation}
\label{eqn:sublatticerotations}
r_1~ =~ \left(0,\frac{1}{6},0,0\right)\;, 
\quad r_2 ~=~ \left(0,0,\frac{1}{3},0\right) 
\quad\text{and}\quad 
r_3 ~=~ \left(0,0,0,\frac{1}{2}\right)\;,
\end{equation}
of order $N=(6,3,2)$, respectively. These rotations act on the world--sheet 
bosons $\boldsymbol{Z} \in \mathbbm{C}^3$ as
\begin{equation}
 \boldsymbol{Z}^i~\xmapsto{\theta^{(i)}}~\mathrm{e}^{2\pi\,\I\,(r_i)^i}\, 
 \boldsymbol{Z}^i \quad\text{for}~i=1,2,3\;.
\end{equation}
Hence, they induce a transformation of the oscillators of \Eqref{eq:twistedstate}, 
i.e.\
\begin{equation}
\label{eqn:trafoexcitations}
\left(\widetilde{\alpha}^{i}_{-\omega_i}\right)^{\widetilde{N}^i}
\left(\widetilde{\alpha}^{\bar{\imath}}_{-1+\omega_i}\right)^{\widetilde{N}^{*i}} 
~\xmapsto{\theta^{(j)}}~ \mathrm{e}^{-2\pi\,\I\, \Delta\widetilde{N}\cdot r_j}
\left(\widetilde{\alpha}^{i}_{-\omega_i}\right)^{\widetilde{N}^i}\, 
\left(\widetilde{\alpha}^{\bar{\imath}}_{-1+\omega_i}\right)^{\widetilde{N}^{*i}}\;,
\end{equation}
where $\Delta \widetilde{N}^i = \widetilde{N}^{*i} - \widetilde{N}^i$ counts the
number of anti--holomorphic ($\widetilde{N}^*$) minus holomorphic 
($\widetilde{N}$) left--moving excitations in the $i^\mathrm{th}$ two--torus. 
The sublattice rotations \eqref{eqn:sublatticerotations} are
accompanied by an  analogous action on the world--sheet fermions of the
right--movers, i.e.\ on $\Ket{q_\text{sh}}_\text{R}$ of \Eqref{eq:twistedstate}.
This action reads
\begin{equation}
\label{eqn:HMomentum}
\Ket{q_\text{sh}}_\text{R} ~\mapsto~ 
\mathrm{e}^{-2\pi\,\I\, q_\text{sh}\cdot r_j}\, \Ket{q_\text{sh}}_\text{R} 
\quad\text{and equivalently}\quad 
\boldsymbol{H} ~\mapsto~ \boldsymbol{H} - \pi\, r_j\;.
\end{equation}
Since $q_\text{sh}^i$ differs by $\nicefrac{1}{2}$ for space--time fermions 
and bosons, these transformations act differently on space--time 
fermions and bosons and hence describe discrete $R$ symmetries in the 
four--dimensional effective theory.

At this step, Kobayashi et al.~\cite{Kobayashi:2004ya} combined the transformation phases 
\eqref{eqn:trafoexcitations} and \eqref{eqn:HMomentum} and defined three $R$ 
charges such that they are invariant under picture changing, i.e.\
\begin{equation}\label{eqn:oldR}
R^{{\text{KRZ}},\,j} ~=~ q_\text{sh}^{j} + \Delta \widetilde{N}^{j}\;.
\end{equation}
For an allowed term in the superpotential these charges have to sum up to $-1$ 
modulo the orders of the sublattice rotation $N^j \in\{6,3,2\}$. 
Note that in this normalization the three $R$ charges \eqref{eqn:oldR} are
fractional, i.e.\ they are multiples of $\nicefrac{1}{6}$, $\nicefrac{1}{6}$
and $\nicefrac{1}{2}$, respectively. In order to normalize them to integers, 
one has to multiply them by $-6$, $-6$ and $-2$. Then the superspace coordinate has $R$
charges $(3,3,1)$ and allowed terms in the superpotential have $R$ charges
$(6,6,2)$ modulo $(36,18,4)$. The orders of the sublattice rotations $N^j$
are different from the orders $M^j$ of the resulting $\Z{M^j}^R$ symmetries,
which are given by
\begin{equation}\label{eqn:ZNR}
\Z{36}^R \times \Z{18}^R \times \Z{4}^R\;.
\end{equation}
However, as first pointed out in \cite{Bizet:2013gf} in the context of orbifolds 
without Wilson lines, also $\Ket{g}$ in \Eqref{eq:twistedstate} transforms in 
general under sublattice rotations. Hence, the $R$ charges \eqref{eqn:oldR} have 
to be amended by contributions from so--called $\gamma$ phases. In the next 
subsection, we present an alternative derivation, which also includes the case 
of non--trivial Wilson lines.

Let us close this subsection with a brief discussion on $T$ moduli. 
The massless spectrum of all Abelian orbifolds contains three diagonal $T$
moduli, denoted by $T_j$ with $j=1,2,3$, associated with the size of the 
$j^\mathrm{th}$ two--torus. The corresponding string states are
\begin{equation}
\label{eqn:TModuli}
T_j ~\sim~ \Ket{q}_\text{R} \otimes \widetilde{\alpha}^{\bar{\jmath}}_{-1} \Ket{0}_\text{L} \otimes \Ket{(\Id,0)}\;,
\end{equation}
with $q_\mathrm{sh}=(0,-1,0,0), (0,0,-1,0), (0,0,0,-1)$ for
$\bar{\jmath}=\bar{1},\bar{2},\bar{3}$. In the effective field theory 
description, the $T$ moduli are chiral superfields. They are gauge singlets 
(since  $p_\text{sh} = 0$) and transform trivially under the space group 
selection rule (since $\Ket{g} = \Ket{(\Id,0)}$). Thus, one can expect 
\Eqref{eqn:oldR} to be the exact form of their $R$ charges which turn out 
to vanish, $R^{{\text{KRZ},\,i}}(T_j) = \delta^i_j - \delta^i_j = 0$. In a 
physical vacuum, the $T_j$ modulus needs to be stabilized at some non--trivial 
value. Hence, (the scalar component of) $T_j$ develops a VEV 
$\langle T_j\rangle$. So we see that the $R$ charges \eqref{eqn:oldR} can 
alternatively be motivated as the unique combination (up to an overall factor) 
of $q_{\text{sh}}$ and $\Delta \widetilde{N}$, such that 
the VEVs of the $T$ moduli do not break the corresponding $R$ symmetries.

\subsection{$\boldsymbol{R}$ charges for twisted fields}
\label{sec:GeometricalEigenstate}

\begin{figure}[!h!]
\centerline{\includegraphics{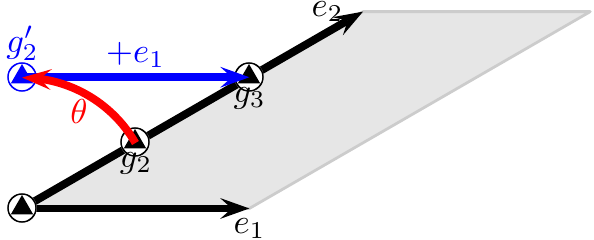}}
\caption{The second twisted sector in the $\text{G}_2$ two--torus has three 
fixed points (black dots) with corresponding constructing elements $g_a$, $a=1,2,3$. Under 
$\theta$ the fixed point of $g_2$ is mapped to $g_2'$, which is equivalent to 
$g_3$ by a lattice translation $+e_1$. Hence, the constructing elements $g_2$ and 
$g_3$ belong to the same conjugacy class.}
\label{fig:Z6fixedpoints}
\end{figure}

As explained above, the geometrical properties of the massless strings are
encoded in $\Ket{g}$, where we identify the fixed point $f_g$ with the
constructing element $g\in\mathbbm{S}$. While $g$ transforms, in general,
non--trivially under the action of $h\in\mathbbm{S}$, 
\begin{equation}
 g~\xmapsto{h}~h\cdot g\cdot h^{-1}=g'\;,
\end{equation} 
the conjugacy class 
\begin{equation}
[g] ~=~ \{ h\cdot g \cdot h^{-1} \;|\; h \in  \mathbbm{S}\}
\end{equation}
is by definition invariant under conjugation. We now construct the corresponding
``geometrical eigenstate'' $\Ket{[g]}$, which is, up to a phase, invariant under
all space--group transformations such that the full physical state
\eqref{eq:twistedstate} is invariant under the action of every
$h\in\mathbbm{S}$. This is achieved by building infinite linear combinations of 
orbifold--equivalent fixed points, or, equivalently, by summing over all
elements of the conjugacy class,
\begin{equation}\label{eq:ket[g]}
\Ket{[g]} ~=~ \sum_h \mathrm{e}^{-2\pi\,\I\, \gamma(g,h)}\, \Ket{h\cdot g \cdot
h^{-1}}\;.
\end{equation}
Here the $\gamma(g,h)$ denote phases that are crucial for rendering
$\Ket{[g]}$ an eigenstate w.r.t.\ all space--group transformations, 
$h\in \mathbbm{S}$ is chosen such that each term 
$\Ket{h\cdot g \cdot h^{-1}}$ appears once in the summation and we suppress the 
normalization. This is a natural extension of the usual linear combination of 
fixed points that are mapped to each other via the twist, e.g.\ in the second 
twisted sector of $\Z{6}$--II the $\text{G}_2$ torus contains three fixed 
points, two of them are identified on the orbifold (cf.\ the discussion in
\cite{Dixon:1986jc,Kobayashi:2004ya,Buchmuller:2006ik}), see 
\Figref{fig:Z6fixedpoints}. However, in contrast to the 
traditional linear combinations, the new geometrical eigenstates are eigenstates of the 
full space group as we will see in more detail later, i.e.\ for any 
$h \in \mathbbm{S}$ one obtains
\begin{equation}
\Ket{[g]} ~\xmapsto{h} ~ \mathrm{e}^{2\pi\,\I\, \gamma(g,h)}\, \Ket{[g]}\;,
\end{equation}
where $\gamma(g,h) \equiv 0$ if $g\cdot h = h\cdot g$. Here and in
what follows ``$\equiv$'' means equal modulo 1. Note that
\eqref{eq:ket[g]} also implies a redefinition of the twist fields 
$\boldsymbol{\sigma}_{g}$, which can be
expressed as an analogous sum. For fixed  $g \in \mathbbm{S}$
the geometrical phase $\gamma(g,h)$ is a homomorphism from the space group
$\mathbbm{S}$ to $\Z{6}$, i.e.\  $\gamma(g,h_1\cdot h_2) \equiv
\gamma(g,h_1) + \gamma(g,h_2)$. Thus, for  $h=(\theta^\ell, m_\alpha\, e_\alpha)$
one has
\begin{equation}
\gamma(g,h)~\equiv~ \ell\, \gamma(g,\theta) + m_\alpha\, \gamma(g,e_\alpha)\;,
\end{equation}
where we define $\gamma(g,\theta) := \gamma\bigl(g,(\theta,0)\bigr)$ and 
$\gamma(g,e_\alpha) := \gamma\bigl(g,(\Id,e_\alpha)\bigr)$. We demand
that the full physical state 
$\Ket{p_\mathrm{sh},q_\mathrm{sh},\widetilde{N},\widetilde{N}^*,g}$
of \Eqref{eq:twistedstate} be invariant under a transformation with $h$. This
translates to the condition
\begin{equation}
\label{eq:orbifoldinvariance}
p_\text{sh}\cdot V_h - \left(q_\text{sh} 
+ \Delta \widetilde{N}\right)\cdot v_h 
- \frac{1}{2}\left(V_g \cdot V_h - v_g \cdot v_h \right) + \gamma(g,h) 
~\stackrel{!}{\equiv}~ 0\;,
\end{equation}
allowing us to compute $\gamma(g,\theta)$ and $\gamma(g,e_\alpha)$ by 
choosing appropriate $h$.

The crucial observation is now that the geometrical eigenstates $\Ket{[g]}$
are eigenstates with respect to a sublattice rotation $\theta^{(j)}$, which
is not an element but an automorphism of $\mathbbm{S}$. Acting with
$\theta^{(j)}$ on $\Ket{[g]}$ yields a phase,
\begin{equation}
\Ket{[g]} ~\xmapsto{\theta^{(j)}} ~ 
\mathrm{e}^{2\pi\,\I\, \gamma(g,\theta^{(j)})}\, \Ket{[g]}\;.
\end{equation}
This is because, as we will show explicitly below, in its action on $\Ket{[g]}$,
$\theta^{(j)}$ is equivalent to an appropriate space--group transformation
$h\in\mathbbm{S}$. In other words, $\theta^{(j)}$ is a conjugacy class
preserving automorphism of $\mathbbm{S}$ (at least for $\Z{6}$--II). Therefore,
the phase $\gamma\left(g,\theta^{(j)}\right)$ can be expressed in terms of
$\gamma(g,\theta)$ and $\gamma(g,e_\alpha)$. 

As we have discussed above \Eqref{eqn:oldR}, sublattice rotations also imply a 
transformation of the shifted \SO8 momenta $q_\mathrm{sh}$ and the oscillator 
numbers $\Delta\widetilde{N}^j$. Taking into account all transformations
under sublattice rotations and using that $(r_j)^j=1/N^j$, the proper 
$R$ charges are thus defined as
\begin{equation}\label{eq:Reff}
R^{j} ~=~ q_\text{sh}^{j} 
       + \Delta \widetilde{N}^{j} 
       - N^{j}\, \gamma(g,\theta^{(j)})\;,
\end{equation}
whose sum must equal $-1$ (modulo the order $N^j$ of the corresponding discrete
sublattice rotation) in order for the correlator~\Eqref{eq:cor} to be
invariant, i.e.\ $\sum_{a=1}^L R^{j}_a = -1 \mod N^j$ for $j=1,2,3$. 
As in \Eqref{eqn:ZNR}, these charges need to be multiplied by 
$(-6,-6,-2)$ in order to make the charges of all fields and of the superspace 
coordinate integer. This charge assignment is valid in the general case 
including non--trivial Wilson lines. In the simplified case without Wilson 
lines it differs by a sign from the previously derived expression in
\cite{Bizet:2013gf}.

In what follows, we discuss this in detail starting with 
sublattice rotations first in the $\SU{3}$ and second in the $\SU{2}^2$ 
two--torus. In these cases, it is sufficient to construct the infinite linear 
combinations for the geometrical eigenstates for each two--torus separately. 
Finally, we perform the sublattice rotation in the $\G2$ two--torus.

\subsubsection{$\boldsymbol{\mathbbm{T}^2_{\SU3}/\Z{3}}$ sublattice rotation}
\label{sec:SU3rotation}

\begin{figure}[!h!]
\begin{center}
\CenterObject{\includegraphics{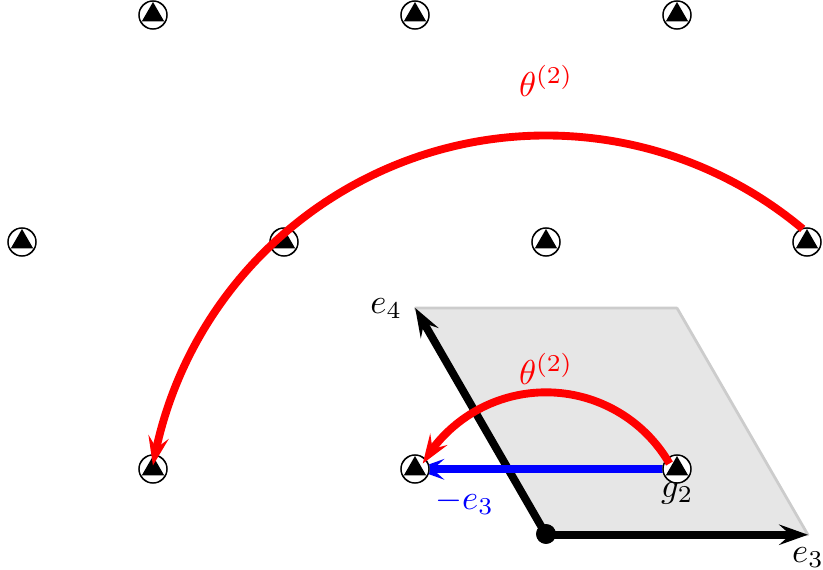}}
\end{center}
\caption{%
Visualization of the geometrical eigenstate $\Ket{[g_2]}$ of
\Eqref{eqn:Z3GeometricalEigenstate}. One performs a sum over all equivalent
fixed points in covering space weighted by appropriate $\gamma$ phase factors.
The sublattice rotation $\theta^{(2)}$ is, for $g_2$, geometrically equivalent
to a lattice translation by $-e_3$ up to three times a lattice translation.}
\label{fig:Z3GeometricalEigenstate}
\end{figure}

Let us consider the second two--torus, where $\Z{6}$--II acts as $\Z{3}$. In 
the first ($k=1$) and fourth ($k=4$) twisted sectors there are three fixed 
points. Their constructing elements read $g_a=(\theta^k, n_3\, e_3 + n_4\, e_4)$,
where $k=1,4$ and $a=1,2,3$ for $(n_3,n_4) = (0,0), (1,0),(1,1)$, 
respectively, see \Figref{fig:SU3Torus}. The associated geometrical
eigenstates $\Ket{[g_a]} $ are obtained by taking infinite linear combinations, 
i.e.\
\begin{equation}
\label{eqn:Z3GeometricalEigenstate}
\Ket{[g_a]} ~=~ \sum_{m_3, m_4} 
\mathrm{e}^{-2\pi\,\I\, (m_3 + m_4)\,\gamma(g_a,e_3)}\, 
\Ket{\bigl(\theta^k, (n_3 + m_3 + m_4)\, e_3 + (n_4 + 2m_4 - m_3)\, e_4\bigr)}\;,
\end{equation}
with $\gamma(g,e_3) \in\{ 0,\nicefrac{1}{3},\nicefrac{2}{3}\}$. Here we sum over
all equivalent fixed points in the covering space, see
\Figref{fig:Z3GeometricalEigenstate}. We can verify  that these three  states
$\Ket{[g_a]}$ are eigenstates of the full space group by letting some 
arbitrary space group element $h$ act on $\Ket{[g_a]}$.
Then each constructing  element $g$ in the linear combination is mapped to
$h\cdot g\cdot h^{-1}$ and,  consequently, $\Ket{[g_a]}$ is mapped to itself
times a phase. For example, under  a general translation $h=(\Id, s_3\, e_3 +
s_4\, e_4)$ the geometrical eigenstate  $\Ket{[g_a]}$ picks up a phase
\begin{equation}
\Ket{[g_a]} ~\xmapsto{h=(\Id, s_3\, e_3 + s_4\, e_4)}~ 
\mathrm{e}^{2\pi\,\I\, (s_3+s_4)\,\gamma(g_a,e_3)}\, \Ket{[g_a]}\;.
\end{equation}
The crucial observation is now that, under a sublattice rotation
$(\theta^{(2)}, 0)$, $\Ket{[g_a]}$ also gets mapped to itself up to a phase, 
\begin{equation}
\Ket{[g_a]}~\xmapsto{(\theta^{(2)}, 0)}~
\mathrm{e}^{-2\pi\,\I\, (n_3 + n_4)\,\gamma(g_a,e_3)}\, \Ket{[g_a]}\;.
\end{equation}
For the case of $\Ket{[g_2]}$, this is illustrated in
\Figref{fig:Z3GeometricalEigenstate}, where we see that any $g_2$--equivalent 
fixed point gets shifted by $-e_3$ up to three times a lattice translation. The 
shift by $-e_3$ induces, in the presence of a Wilson line, a non--trivial phase 
while, due to the Wilson line quantization and modular invariance conditions, 
three times a lattice translation does not lead to a phase. Thus, we find that
$\gamma(g_a,\theta^{(2)}) = -(n_3 + n_4)\,\gamma(g_a,e_3)$ is  the contribution
to the $R$ charge under a rotation  $r_2 = (0,0,\frac{1}{3},0)$ for a state from
the $k=1,4$ sectors. For a state  from the $k=2,5$ sector we get
$\gamma(g_a,\theta^{(2)}) = (n_3 + n_4)\,\gamma(g_a,e_3)$.  Finally, for $k=0,3$
we have $\gamma(g_a,\theta^{(2)}) = 0$. Combining these  results we obtain
\begin{equation}
\gamma(g_a,\theta^{(2)}) ~\equiv~ -k\, (n_3 + n_4)\,\gamma(g_a,e_3) 
\end{equation}
for a state with constructing element $g_a = (\theta^k, n_3\, e_3 + n_4\, e_4)$.

Altogether we have seen that the sublattice rotation $\theta^{(2)}$, whose gauge
embedding is not defined (because $\theta^{(2)}\not\in\mathbbm{S}$), can be
traded against a translation, for which we know the gauge embedding. From this
we can infer the transformation properties of the full state
$\Ket{p_\mathrm{sh},q_\mathrm{sh},\widetilde{N},\widetilde{N}^*,g}$
(see~\Eqref{eq:twistedstate}). Demanding that
$\Ket{p_\mathrm{sh},q_\mathrm{sh},\widetilde{N},\widetilde{N}^*,g}$ be invariant
under all space group transformations allowed us then to compute  via
\Eqref{eq:orbifoldinvariance} the $\gamma$ phases, which enter the  $R$ charges
\eqref{eq:Reff}.

\subsubsection{$\boldsymbol{\mathbbm{T}^2_{\SU2\times\SU2}/\Z{2}}$ sublattice rotation}
\label{sec:SO4rotation}

Next, consider the two--torus $\mathbbm{T}^2_{\SU2\times\SU2}$, where
$\theta$ acts as $\Z{2}$. The analysis is analogous to the one above. In this 
torus there are four fixed points (if $k$ is odd) with constructing elements
\begin{equation}
g_a~=~(\theta^k, n_5\, e_5 + n_6\, e_6)\;,
\end{equation}
where $(n_5,n_6) = (0,0), (0,1), (1,0)$ or $(1,1)$ for $a=1,2,3,4$, respectively, 
see \Figref{fig:SO4Torus}. 
Again, the associated geometrical eigenstates $\Ket{[g_a]} $ are obtained by taking 
infinite linear combinations, i.e.
\begin{equation}
\Ket{[g_a]}~=~\sum_{m_5, m_6} 
\mathrm{e}^{-2\pi\,\I\, \bigl(m_5\,\gamma(g_a,e_5) + m_6\,\gamma(g_a,e_6)\bigr)}\, 
\Ket{\bigl(\theta^k, (n_5 + 2m_5)\, e_5 + (n_6 + 2m_6)\, e_6\bigr)}\;,
\end{equation}
with $\gamma(g_a,e_5)\in\{ 0,\nicefrac{1}{2}\}$ and $\gamma(g_a,e_6) \in\{
0,\nicefrac{1}{2}\}$. As before, under a general translation $h=(\Id, s_5 e_5 +
s_6 e_6)$ the geometrical eigenstate $\Ket{[g_a]}$ picks up a phase
\begin{equation}
\Ket{[g_a]}~\mapsto~
\mathrm{e}^{2\pi\,\I\, \bigl(s_5\,\gamma(g_a,e_5) + s_6\,\gamma(g_a,e_6)\bigr)}
\, \Ket{[g_a]}\;.
\end{equation}
Furthermore, under a sublattice rotation $(\theta^{(3)}, 0)$, $\Ket{[g_a]}$
transforms with a phase,
\begin{equation}
\Ket{[g_a]}~\mapsto~
\mathrm{e}^{2\pi\,\I\,\bigl(n_5\gamma(g_a,e_5) + n_6\gamma(g_a,e_6)\bigr)}\, 
\Ket{[g_a]}\;.
\end{equation}
If $k$ is even, the sublattice rotation $\theta^{(3)}$ acts on a fixed torus 
(with $n_5 = n_6 = 0$) and hence $\gamma(g_a,\theta^{(3)}) = 0$. Combining 
these results we obtain
\begin{equation}
\gamma(g_a,\theta^{(3)}) ~\equiv~ n_5\,\gamma(g_a,e_5) + n_6\,\gamma(g_a,e_6) \;,
\end{equation}
for a state with constructing element $g_a = (\theta^k, n_5\, e_5 + n_6\, e_6)$.

Let us stress that our analysis in \ref{sec:SU3rotation} and
\ref{sec:SO4rotation} can be applied to any orbifold with a $\Z{3}$ 
sublattice rotation in an \SU3 plane and/or $\Z{2}$ sublattice rotation in an 
$\SU2\times\SU2$ plane, thus allowing us to compute the proper $R$ charges for
many other orbifold geometries.

\subsubsection{$\boldsymbol{\mathbbm{T}^2_{\G2}/\Z{6}}$ sublattice rotation}

Last, we consider the first complex plane, where $\Z{6}$--II acts as $\Z{6}$. 
There are two ways to derive $\gamma(g,\theta^{(1)})$ in this case. First, 
we know that
\begin{equation}
\theta^{(1)} ~=~ \theta\cdot \left(\theta^{(2)}\right)^{-1}\cdot 
\left(\theta^{(3)}\right)^{-1}\;.
\end{equation}
Hence, the phase of the geometrical eigenstate with constructing element 
$g = (\theta^k, n_\alpha e_\alpha)$ under a $\theta^{(1)}$ 
sublattice rotation is given by
\begin{eqnarray}
\gamma(g,\theta^{(1)}) & \equiv & \gamma(g,\theta) - \gamma(g,\theta^{(2)}) - \gamma(g,\theta^{(3)}) \\
                       & \equiv & \gamma(g,\theta) + k\, (n_3 + n_4)\,
				   \gamma(g,e_3) - \bigl(n_5\,\gamma(g,e_5) +
				   n_6\,\gamma(g,e_6)\bigr)\;.
\end{eqnarray}
The second possibility is the explicit construction of the full  geometrical
eigenstate, which yields the same result for $\gamma(g,\theta^{(1)})$, as
expected.

\subsubsection{Summary of $\boldsymbol{R}$ charges}

In summary, the three $R$ charges of the $\Z{36}^R\times\Z{18}^{R}\times\Z4^R$ 
symmetry for a (twisted) state of the $\Z{6}$--II orbifold with constructing 
element $g = \left(\theta^k, n_\alpha\, e_\alpha\right)$ read
\begin{subequations}
\begin{eqnarray}
R^1 & = & -6\, \Bigl[q_\text{sh}^1 + \Delta \widetilde{N}^1 
          -6\,\gamma(g,\theta)\nonumber \\
    &   &{} \hphantom{-6\, \Bigl[}{}  
          -6\, k\, (n_3 + n_4)\,\gamma(g,e_3)
          +6\,\bigl(n_5\,\gamma(g,e_5) + n_6\,\gamma(g,e_6)\bigr)\Bigr]\;, \\
R^2 & = & -6\, \Bigl[q_\text{sh}^2 + \Delta \widetilde{N}^2 + 3\, k\, (n_3 + n_4)\,\gamma(g,e_3)\Bigr]\;, \\
R^3 & = & -2\, \Bigl[q_\text{sh}^3 + \Delta \widetilde{N}^3 -
2\,\bigl(n_5\,\gamma(g,e_5) + n_6\,\gamma(g,e_6)\bigr)\Bigr]
\;,
\end{eqnarray}
\end{subequations}
where the superspace coordinate has $R$ charges $(3,3,1)$ and all $R$ charges 
are normalized to be integer. Note that the $\gamma$ charges vanish for 
untwisted fields. We have ``tested'' these $R$ charges for a huge set of 
randomly generated \Z6--II orbifold models with non--trivial Wilson lines 
\cite{Nilles:2011aj} and found that the anomalies are always universal, i.e.\ 
can be cancelled by the dilaton. On the other hand, restricting to $W_\alpha=0$ 
and using the $R$ charges from \cite{Bizet:2013gf}, where the 
$N^j\, \gamma(g,\theta^{(j)})$ term appears with the opposite sign, leads to 
non--universal $R^{1}$ anomalies.

It is instructive to apply the three discrete $R$ transformations
consecutively to some field $\Psi_g$. This results in a $\Z{36}$ phase $R_g$
given by
\begin{equation}
\frac{1}{36} R_g~=~ -\frac{1}{36}
\left(R^1 + 2 R^2 - 9 R^3\right) 
~\equiv~ p_\text{sh}\cdot V - \frac{1}{2}\left(V_g \cdot V - v_g \cdot v \right)\;,
\end{equation}
where we used the invariance condition \Eqref{eq:orbifoldinvariance}. Now
consider a coupling between states with constructing elements  $g_1 \ldots
g_L$. One can see that the total $R$ transformation is trivial, i.e.\
$\frac{1}{36} \sum_g R_g \equiv 0$ by using gauge invariance, the point
group selection rule, the space group selection rule in the second and third
two--torus and finally modular invariance. Hence, the string selection rules
are not independent and one could trade off, for example, one of the $R$
symmetries. That is, as also observed in \cite{Buchmuller:2008uq}, some of the 
symmetries are redundant; these redundancies can be eliminated with the methods 
discussed in \cite{Petersen:2009ip}.

One may also wonder if one could separate off the $\gamma$ contributions
from the $R$ charges. At first glance, one may think the space--group selection
rule implies that the $\gamma$--phases sum up to $0\mod1$ since the product of
the respective constructing elements has to yield the identity
$(\Id,0)$, and $\gamma\bigl((\Id,0),h\bigr)\equiv0$ for all $h\in\mathbbm{S}$.
However, for each constructing element $g\in\mathbbm{S}$ the sublattice
rotations $\theta^{(i)}$ are, in general, equivalent to \emph{different}
space--group operations such that it is not generally possible to separate the
$\gamma$ contributions.


\section{Summary}
\label{sec:summary}

We have re--derived the $R$ symmetries and charges for the \Z6--II orbifold with
Wilson lines. 
As we have seen, the discrete $R$ symmetries originate from sublattice rotations 
of the orbifold accompanied by an analogous action on the right--mover. This 
yields the well--known contributions to the $R$ charges. By constructing
states that are invariant under the full space group $\mathbbm{S}$ we were able
to determine the transformation behavior of the twist fields under sublattice
rotations, which are automorphisms but not elements of $\mathbbm{S}$. 
Separating the correlator of the vertex operators into a gauge part and a 
rest allowed us to determine necessary conditions for the correlators to be
non--trivial, which can be rewritten as discrete $R$ symmetries. With our
derivation, we confirm the statement of \cite{Bizet:2013gf} that the $R$ charges
have to be amended by appropriate $\gamma$ phases, disagree, however, in a sign.
Further, our derivation allowed us to treat also the case of non--vanishing
Wilson lines. 

Using the correct definition of $R$ charges, \Eqref{eq:Reff}, has
important  consequences for orbifold model building.  First, $\mathbbm{Z}_M^R$
anomalies are now universal,  as we have explicitly verified in thousands of
\Z6--II orbifold models  (including up to three non--trivial Wilson lines).  In
particular, the non--universal anomalies found in \cite{Araki:2008ek} are a
consequence of the incorrect $R$ charges used in the analysis. Repeating the
analysis with proper $R$ charges leads to universal anomalies, which can be
cancelled by the dilaton. Further, the fact that \cite{Araki:2007ss} did find
universal  anomalies ignoring the $\gamma$ phases is, in particular, related to
the simplicity of their  models which is characterized by the absence of Wilson
lines, such that the massless  twisted states appear with degeneracy factors,
thus rendering the anomaly coefficients universal ``by accident''. Using
proper $R$ charges has also important implications for heterotic orbifold
phenomenology. In particular, if one compares couplings that are allowed by the
incorrect vs.\ correct $R$ charges, one finds that  many more couplings are
allowed if one imposes the proper $R$ symmetries.  As a consequence,
vector--like exotics of MSSM--like constructions, such as those
of~\cite{Lebedev:2006kn,Lebedev:2007hv,Kappl:2008ie}, decouple at low orders
and Yukawa textures are changed.  At the same time, discrete $R$ symmetries
(such as the $\Z4^R$ symmetry~\cite{Lee:2010gv}) remain instrumental for
suppressing the $\mu$ term and dangerous proton decay operators. Yet, clearly,
the construction of vacua with residual discrete and/or approximate $R$
symmetries has to be revisited. This will be done elsewhere.

Although our presentation was focused on the \Z6--II orbifold based on
factorizable tori, our derivation is general and can be extended to all
symmetric (or geometric) orbifolds \cite{Fischer:2012qj}.  In particular, our
analysis in sections~\ref{sec:SU3rotation} and \ref{sec:SO4rotation} can be
applied to any orbifold with a $\Z{3}$  sublattice rotation in an \SU3 plane
and/or $\Z{2}$ sublattice rotation in an  $\SU2\times\SU2$ plane, thus allowing
us to compute the proper $R$ charges for many other orbifold geometries. This
analysis will be carried out elsewhere.

\paragraph{Acknowledgements.} We would like to thank Robert Richter for 
important discussions. S.R--S.\ thanks the ICTP for hospitality and the 
support received through the ICTP Junior Associateship Scheme. M.R.\ would like
to thank the  UC Irvine, where part of this work was done, for  hospitality. 
The work of H.P.N.\ was supported by SFB-Transregio TR33 ``The Dark Universe''
(Deutsche Forschungsgemeinschaft) and the European Union 7th network program
``Unification in the LHC era'' (PITN-GA-2009-237920). This work was partially
supported by the DFG cluster of excellence ``Origin and Structure of the
Universe''. P.V.\ is supported by SFB grant 676.  This research was done in the
context of the ERC  Advanced Grant project ``FLAVOUR''~(267104). S.R--S.\ is
partially supported by CONACyT grant 151234 and DGAPA-PAPIIT grant
IB101012-RR181012.

\bibliography{Orbifold}
\bibliographystyle{NewArXiv}

\end{document}